\input{aipcheck}

\documentclass[
    ,final            
  ]
  {aipproc}

\layoutstyle{6x9}
\usepackage{amsmath,amssymb,amsfonts,amscd,bbm,cancel,graphicx}

\begin{document}

\title{Discretized Gravity on the Hyperbolic Disk\footnote{Talk
  presented at SUSY06, the 14th Conference on Supersymmetry and the Unification of Fundamental
Interactions, UC Irvine, California, 12-17 June 2006}}

\author{Gerhart Seidl\footnote{E-mail:\texttt{seidl@physik.uni-wuerzburg.de}}}{
  address={Institut f\"ur Theoretische Physik und Astrophysik\\
Universit\"at W\"urzburg, Am Hubland\\
97074 W\"urzburg, Germany}}

\begin{abstract}
We consider discretized gravity in six dimensions, where the two extra
dimensions have been compactified on a hyperbolic disk of constant
curvature. We analyze a fine-grained realization of lattice gravity on the
hyperbolic disk at the level of an effective field theory for massive
gravitons. It is shown that a nonzero curvature or warping in radial
direction allows to obtain a strong coupling scale that becomes in the
infrared regime larger than in discretized warped five-dimensional space. In
particular, when approaching the boundary of the discretized warped
hyperbolic disk, the local strong coupling scale can be as
large as the local Planck scale.
\end{abstract}

\maketitle


\section{Introduction}
Curved space-time manifolds admit to realize a number of extremely
interesting ideas ranging from the Randall-Sundrum (RS) models
\cite{Randall:1999ee} to the AdS/CFT
correspondence \cite{Maldacena:1997re} or moduli stabilization in
string compactifications with fluxes \cite{Grana:2005jc}. A strong
space-time curvature is also advantageous when formulating lattice
gravity in the context of an effective field theory
(EFT) \cite{Arkani-Hamed:2002sp,Arkani-Hamed:2003vb}. In 5D flat
space, the ultraviolet (UV) strong
coupling scale depends on the bulk volume, or infrared (IR) scale, via a
so-called ``UV/IR connection'' that would forbid to take the large volume limit
within a sensible EFT \cite{Arkani-Hamed:2003vb}. Recently, it has
been shown that this UV/IR
connection can be avoided for discretized gravity in five-dimensional (5D)
warped space-time \cite{Randall:2005me,Gallicchio:2005mh} such that a
large volume limit like in RS II becomes possible. However, the strong coupling scale is in
discretized 5D warped space still everywhere smaller than the local Planck scale.

In this talk, we consider a six-dimensional (6D) lattice gravity construction, where the two extra
dimensions are compactified on a discretized hyperbolic disk
\cite{Bauer:2006pf} (see also Ref.~\cite{Bauer:2006ti}).
We study a fine-grained latticization of a hyperbolic disk that is
warped along the radial direction and estimate there the local strong coupling scale. It turns out that the
presence of the 6th dimension can yield the theory on the discretized warped
hyperbolic disk more weakly coupled than in the corresponding
5D warped case. In particular, the local strong coupling scale can
become on the boundary of the disk as large as the local Planck scale.

\section{6D warped hyperbolic space}\label{sec:continuum}
\begin{figure}
\includegraphics[bb = 220 615 390 750, height=.2\textheight]{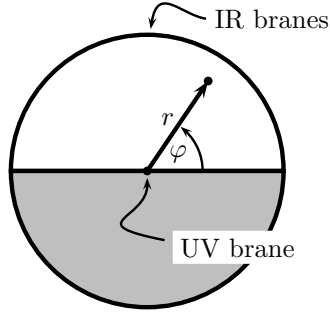}
\caption{Physical space of the orbifold $K_2/Z_2$. Shown are the
    coordinates $(r,\varphi)$ of a point on the hyperbolic disk $K_2$,
    which is warped along the radial direction. The orbifold projection $\varphi\rightarrow-\varphi$
    identifies the upper half of $K_2$ with the lower half (gray
    shaded region).}
\label{fig:orbifold}
\end{figure}

Consider 6D general relativity compactified to four dimensions
on an orbifold $K_2/Z_2$, where $K_2$ is a two-dimensional hyperbolic disk of constant
negative curvature. The 6D coordinates $x^M$ are labeled by capital
Roman indices $M=0,1,2,3,5,6$ while Greek indices $\mu=0,1,2,3$ denote
the usual 4D coordinates $x^{\mu}$. The
6D Minkowski metric is $\eta_{MN}=\textrm{diag}(1,-1,\dots,-1)$. A point on the hyperbolic disk $K_2$ is described by polar coordinates $(r,\varphi)$, where
$r=x^{5}$ and $\varphi=x^{6}$ with $r\in[0,L]$ and $\varphi\in[0,2\pi)$. Here, $r$ is
the geodesic distance of the point $(r,\varphi)$ from the
center, {\it i.e.}, $L$ is the hyperbolic radius of the disk. 
The disk is warped along the $r$ direction and the 6D metric
for the warped hyperbolic disk is given by the line element
\begin{equation}
\textrm{d}s^{2}=e^{2\sigma(r)}g_{\mu\nu}(x^\mu,r,\varphi)\textrm{d}x^{\mu}\textrm{d}x^{\nu}-\textrm{d}r^{2}-v^{-2}\,
\textrm{sinh}^2(vr)\textrm{d}\varphi^{2},\label{eq:6D-metric}\end{equation}
 where $1/v>0$ is the curvature radius of the disk,
 $g_{\mu\nu}(x^\mu,r,\varphi)$ is the 4D metric, and
$\sigma(r)=-wr$, where $w$ is the curvature scale for the
 warping. The geometry of the orbifold $K_2/Z_2$ with the definition of
 the UV and IR branes is shown in Fig.~\ref{fig:orbifold}.
Expanding $g_{\mu\nu}$ in terms of small fluctuations around 4D
Minkowski space as $g_{\mu\nu}=\eta_{\mu\nu}+h_{\mu\nu}$, the relevant 6D kinetic part of the graviton
Lagrangian density is then, to quadratic order, found to be of the form
\begin{eqnarray}\label{eq:linearized}
\mathcal{S}_\textrm{lin}&=& M_6^4\int\textrm{d}^6x\,v^{-1}
\textrm{sinh}(vr)\Big[\frac{1}{4}e^{4\sigma(r)}
(\partial_{r}h_{\mu\nu})(\eta^{\mu\nu}\eta^{\alpha\beta}-\eta^{\mu\alpha}\eta^{\nu\beta})(\partial_rh_{\alpha\beta})\nonumber\\
&&\qquad\,\,+\frac{1}{4}e^{4\sigma(r)}v^2\textrm{sinh}^{-2}(vr)
(\partial_\varphi
h_{\mu\nu})(\eta^{\mu\nu}\eta^{\alpha\beta}-\eta^{\mu\alpha}\eta^{\nu\beta})(\partial_\varphi
h_{\alpha\beta})\Big],
\end{eqnarray}
where $M_{6}$ is the 6D Planck scale, $h={h^{\mu}}_{\mu}$, $h_{\nu}=\partial^{\mu}h_{\mu\nu}$, and $h_{5M}=h_{6M}=0$.

\section{Discretization and Strong Coupling}\label{sec:refined}
\begin{figure}
\includegraphics*[bb = 200 520 400 700, height=.25\textheight]{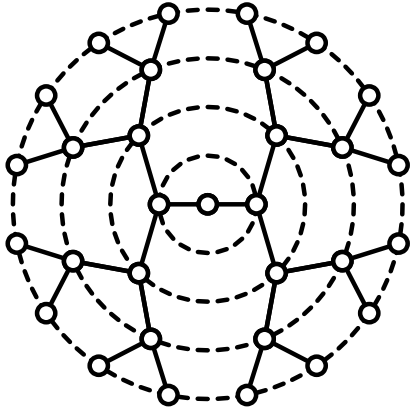}
\caption{\small Fine-grained discretization of the hyperbolic disk for
$k_{\rm max}=4$. The small solid circles represent the
    sites and are labeled by an index $i$. On each site $i$ lives a
    graviton field $g_{\mu\nu}^i$ and two neighboring sites $i$
    and $j$ on adjacent circles are connected by a link $(i,j)$ (solid lines).
}\label{fig:tessellation}
\end{figure}

Let us consider a fine-grained latticization of the hyperbolic
disk $K_2$ which is of the type shown in Fig.~\ref{fig:tessellation},
where we will throughout interpret the
sites (drawn as circles) and links (solid straight lines) following
Ref.~\cite{Arkani-Hamed:2002sp}.
The sites are distinguished by an index $i$ and each site $i$ is equipped with its own metric
$g^{i}_{\mu\nu}$ that can be expanded around 4D flat space as
$g^i_{\mu\nu}=\eta_{\mu\nu}+h^i_{\mu\nu}$. The sites lie on
$k_\textrm{max}$ concentric circles (dashed lines) that are labeled
as $k=0,1,2,\dots,k_{\rm max}$ when going from the center ($k=0$) to
the boundary ($k=k_{\rm max}$). Fig.~\ref{fig:tessellation} shows the
special case $k_{\rm max}=4$.
On the $k$th circle, we have $N_k\equiv 2^k$ sites, and the
  total number of sites on the first $k$
circles is $N_k^{\rm total}\equiv 2^{k+1}-1$. The exponential growth
of $N_k$, when moving outward on the disk, is
a salient feature of our graph that is important for the strong coupling
behavior near the IR branes on the
boundary. Like in Ref.~\cite{Randall:2005me}, we assume that the radial geodesic coordinates of the sites
are integer multiples of a common proper radial lattice spacing
  $a\equiv 1/m_*$ and take
the values $k\cdot a$, where $k=0,1,2,\dots, k_\textrm{max}$, {\it i.e.}, the $k$th
circle has a proper radius $k\cdot a$. For a given site $i$, we will
  denote by $r_i$ the proper radius of the concentric circle on
  which the site $i$ is located.  To introduce the warping along the radial direction of the fine-grained latticized disk, we proceed exactly like in the discretized 5D RS model
in Ref.~\cite{Randall:2005me} and replace in the
linearized gravitational action in Eq.~(\ref{eq:linearized}), {\it e.g.}, the
derivatives in the $r$ direction by $e^{-2wr}\partial_r
h_{\mu\nu}\rightarrow\frac{1}{a}e^{-2wr_i}
(h_{\mu\nu}^{j}-h_{\mu\nu}^{i})$,
where it is understood that $i$ and $j$ are neighboring sites on two
adjacent concentric circles that are connected by a link $(i,j)$.
We thus obtain for the discretized gravitational Lagrangian on the warped
hyperbolic disk $\mathcal{L}_\textrm{lin}=M_4^2
\sum_ie^{-2wr_i}e^{-2wr_i}h^{i}_{\mu\nu}\Box
 h^{i}_{\mu\nu}+\mathcal{L}_\textrm{FP}$,
where $\mathcal{L}_\textrm{FP}$ are Fierz-Pauli mass terms that
are schematically given by
\begin{equation}\label{eq:warpedFP}
\mathcal{L}_{\textrm{FP}}=M_{4}^{2}\sum_{(i,j)}
m_{*}^{2}e^{-4wr_i}(h_{\mu\nu}^{j}-h_{\mu\nu}^{i})(\eta^{\mu\nu}\eta^{\alpha\beta}-\eta^{\mu\alpha}\eta^{\nu\beta})(h_{\alpha\beta}^{j}-h_{\alpha\beta}^{i}),
\end{equation}
where $i$ and $j$ are neighboring sites connected by a link
$(i,j)$ and $M_4$ is the universal 4D Planck scale on the
sites. Restoring like in
Refs.~\cite{Arkani-Hamed:2002sp,Randall:2005me} general coordinate invariances by adding Goldstone bosons
to $\mathcal{L}_{\rm lin}$, we find for the local
strong coupling scale on the $k$th circle \cite{Bauer:2006pf}
\begin{equation}\label{eq:localscale}
 \Lambda_\textrm{warp}^{6D}(k)=\sqrt{\frac{w}{M_4}}(\sqrt{N_{k}}M_k
 m_\ast^4e^{-4wka})^{1/5}=N_k^{1/10}\Lambda^{5D}_{\rm warp}(k),
\end{equation}
where $M_k\equiv M_4e^{-wka}$ is the local Planck scale on the $k$th circle and
$\Lambda^{5D}_{\rm warp}(k)$ is the strong coupling scale in
the corresponding 5D warped case. The important point is here the
presence of the factor $N_k$ that
can render the model more weakly coupled than in 5D warped space when
$N_k$ becomes exponentially large. For $w/M_4\simeq 0.1$, we see that
having $\Lambda_{\rm warp}^{6D}(k)$ as large as $M_k$ requires
exponentially many sites $N_k$ on the $k$th circle, which is in our
model possible in the outer regions of $K_2$ at the IR branes. In 5D warped space, $\Lambda^{5D}_{\rm warp}(k)$ is always smaller than the
local Planck scale. Choosing, instead, in our 6D model,
{\it e.g.}, $m_\ast=M_4$ and $w=(0.1)\times M_4$, the local strong
coupling scale can become on the hyperbolic disk as large as the local
Planck scale $M_k$ for values $k=\mathcal{O}(10^2)$.

To summarize, we have seen for two discrete
gravitational extra dimensions compactified on a warped hyperbolic disk that
a high curvature or strong warping allows to avoid the UV/IR connection problem of
 lattice gravity in flat space. By going to six dimensions on the
 hyperbolic disk, it is possible to improve further the strong coupling behavior of the
 5D flat or warped case and achieve on the IR branes a description of
 lattice gravity that is valid up to the local Planck scale. It would be interesting to relate our analysis, {\it
 e.g.},  also to moduli stabilization in
 effective theories, to theories with
 spontaneously broken space-time symmetries \cite{Kirsch}, and to studies
 on the implications of latticized extra dimensions for cosmology \cite{Hallgren:2005mw}.


\begin{theacknowledgments}
  I would like to thank my collaborators Florian Bauer and Tomas
  H\"allgren. This work was supported by the U.S. Department of Energy
  under grant number DE-FG02-04ER46140.
\end{theacknowledgments}

\end{document}